\documentclass[a4paper, amsfonts, amssymb, amsmath, reprint, showkeys, nofootinbib, twoside]{revtex4-1}
\usepackage[english]{babel}
\usepackage[utf8]{inputenc}
\usepackage{ulem}
\usepackage[colorinlistoftodos, color=green!40, prependcaption]{todonotes}
\begin{document}
\title{Earthquake magnitude distribution and aftershocks: a statistical geometry explanation}

    
\author{Fran\c{c}ois P\'etr\'elis} 
\affiliation{Laboratoire de Physique de l'Ecole normale supérieure, ENS, Université
PSL, CNRS, Sorbonne Université, Université Paris-Diderot - Paris, France}
\author{Kristel Chanard}
\affiliation{Université de Paris, Institut de physique du globe de Paris, CNRS, IGN, F-75005 Paris, France }
\author{Alexandre Schubnel}
\affiliation{Laboratoire de G\'eologie, CNRS UMR 8538, Ecole normale Sup\'erieure, PSL Research University, Paris, France}
\author{Takahiro Hatano}
\affiliation{Department of Earth and Space Science, Osaka University, 560-0043 Osaka, Japan}

\date{\today} 

\begin{abstract}
The emergence of a power-law distribution for the energy released during an earthquake is investigated in several models. Generic features are identified which are based on the self-affine behavior of the stress field prior to an event. This field behaves at large scale as a random trajectory in 1 dimension of space and a random surface in 2 dimensions. Using concepts of statistical mechanics and  results on the properties of these random objects, several predictions are obtained and verified, in particular the value of the power-law exponent of the earthquake energy distribution (the Gutenberg-Richter law) as well as a mechanism for the existence of aftershocks after a large earthquake (the Omori law). 
 
\end{abstract}


\maketitle

Two of the most widely observed and intriguing properties of earthquakes (EQ) are the distribution of energy that they release and the variation with time of the number of aftershocks following a mainshock. 
Using modern definitions,  the released energy during an EQ is characterized by the magnitude $m$,  $m=2 \log_{10}(M)/3$ where $M$ is the moment $M\propto \sum  \Delta x$. The sum is taken over all the spatial extent that has moved during the EQ  and $ \Delta x$ is the total displacement during the event.  

In natural data,  the distribution of $m$ is observed to be an exponential, so-called Gutenberg-Richter (GR) law \cite{bookEQ}. It is written $P(m) \propto 10^{-b\, m}$ and the value of $b$ usually ranges between $3/4$ and $1$ \cite{valb}. Translated into the distribution of the moment $M$, the GR law turns into a power law  $P(M)\propto M^{-1-B}$, where $B=2 b /3$ and  ranges between $1/2$ and $2/3$ \cite{Takareview}. For what concerns  the number of aftershocks per unit time,  Omori's law \cite{Omori} states that $dn/dt\propto (t+\tau)^{-a}$ where  $a$ is of order unity, $\tau$ is a constant and $t$ is the duration since the mainshock. 

There exists a variety of EQ models that use simplified  dynamical rules to describe the evolution of faults, see for instance \cite{BK,OFC,Takareview,PRE77}. 
Here, from the study of several of such models, we identify an analogy between the nonlinear dynamics of the solutions of EQ models and some statistical properties of random curves or surfaces.  
Using methods and tools of statistical mechanics, we are able to explain the origin of the GR law and the value of its exponent.  In addition, in 2 dimensions (2D) this approach provides an explanation for the existence of aftershocks after a large earthquake.

\begin{figure*}
\centerline{a\includegraphics[width=9cm]{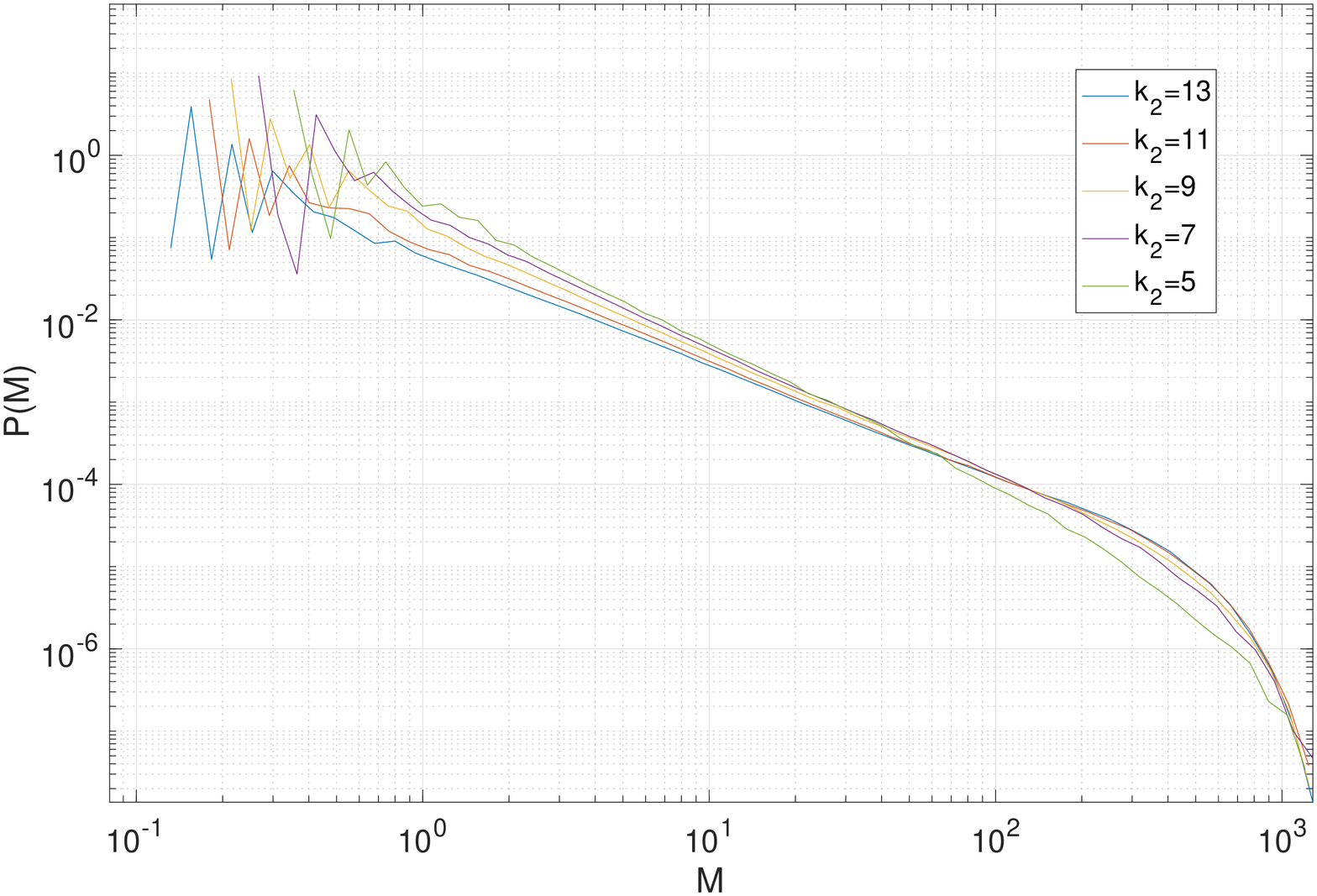}
b\includegraphics[width=9cm]{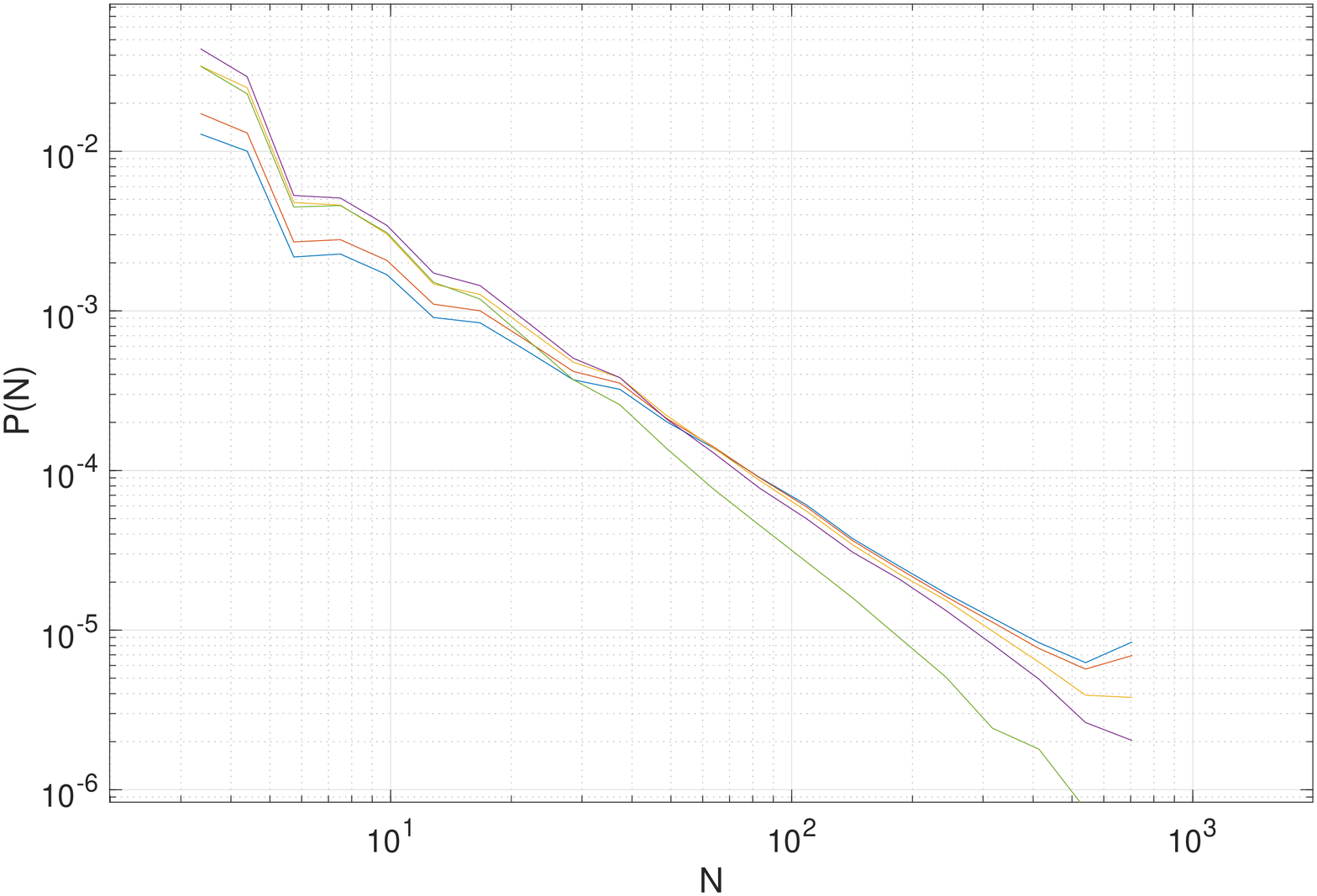}}
\centerline{c\includegraphics[width=9cm]{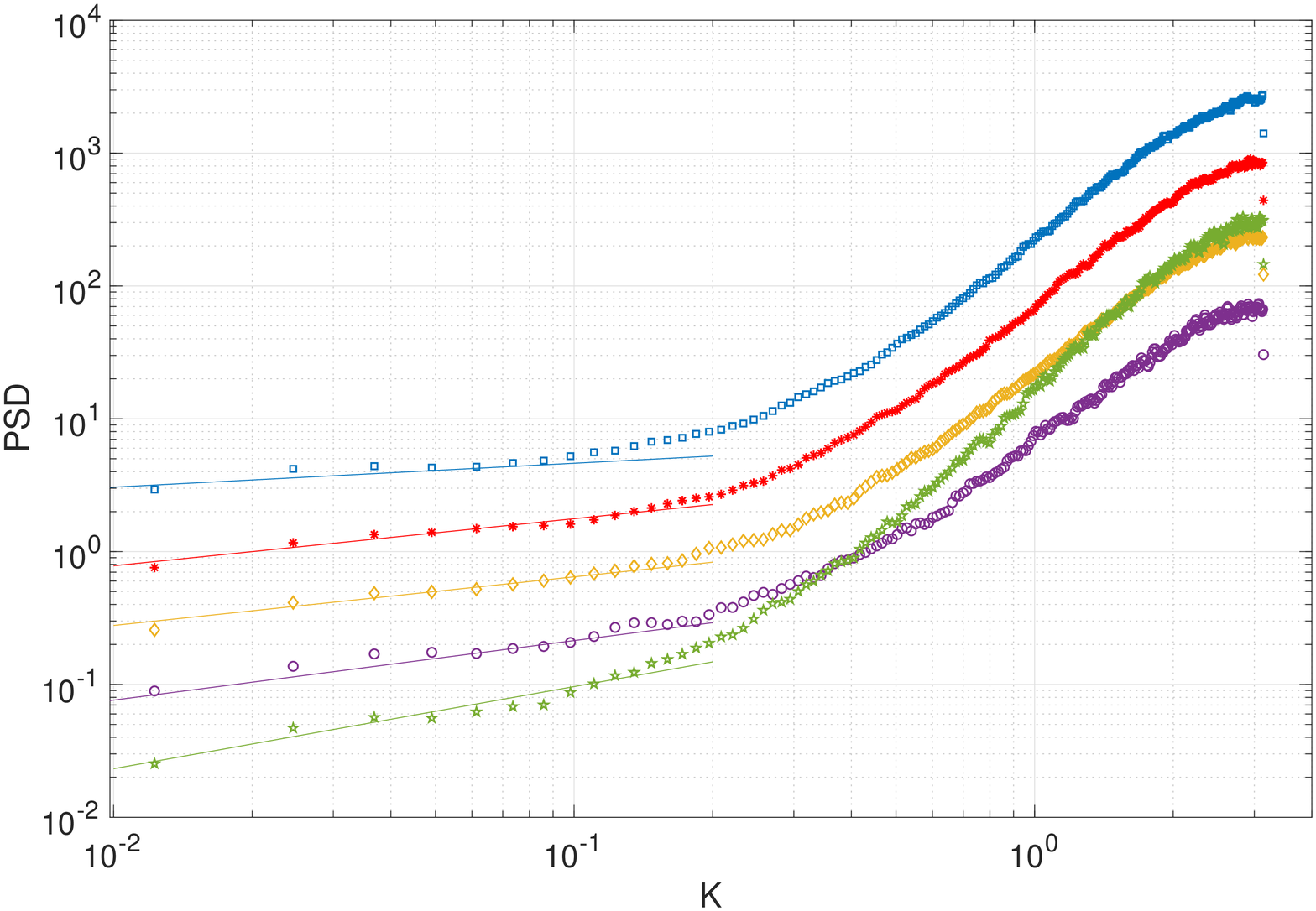}
d\includegraphics[width=9cm]{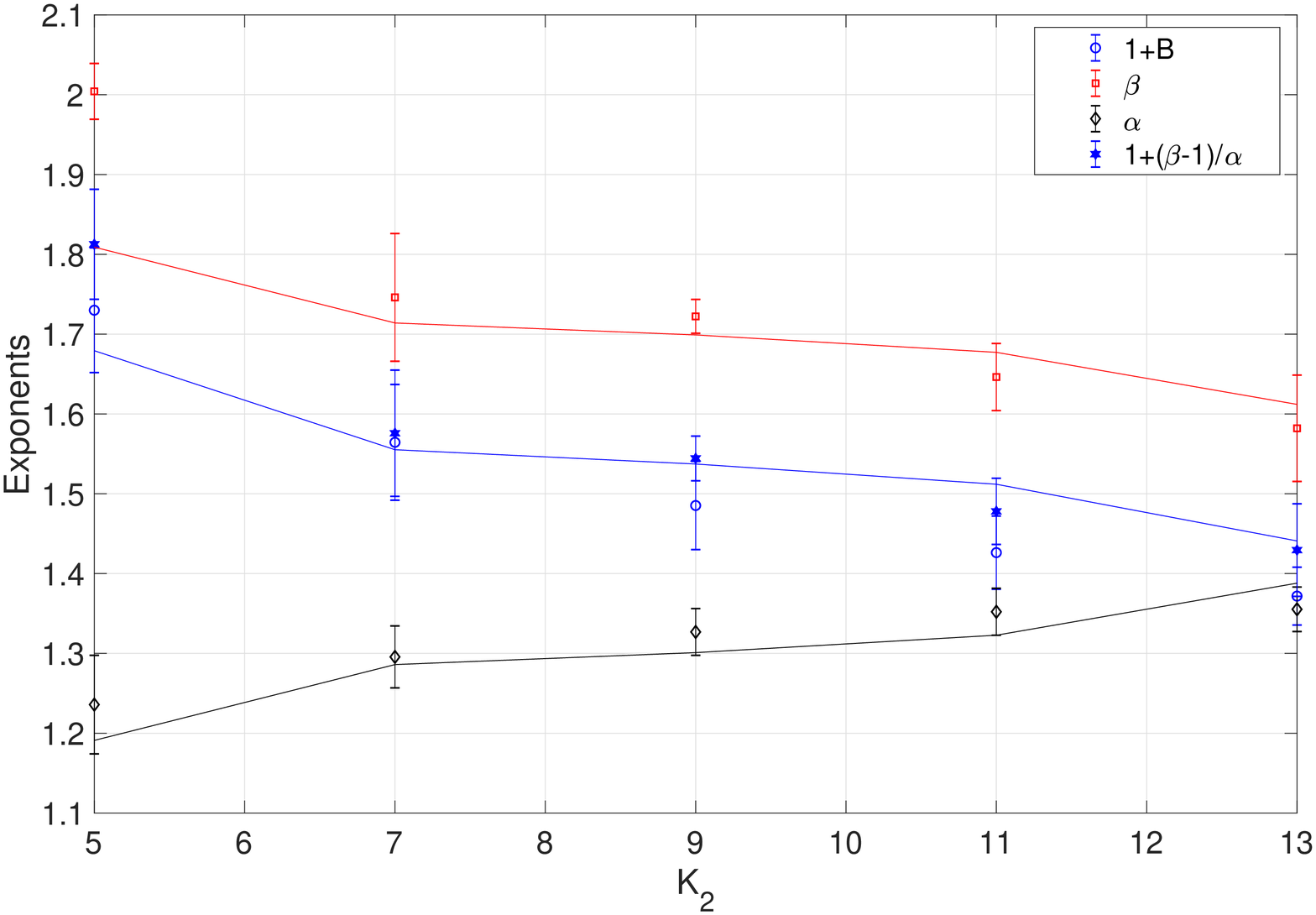}}
\caption{For the solution of the Coulomb  BK model in 1D with $F_d=0$, $N=800$, $v_0=10^{-6}$, $k_1=1$ and varying $k_2$ (see color code in the legend), a: Probability density function (PDF) of the moment of the events $M$.
b: PDF of the event size $N$.
c: Power spectrum density (PSD) of the spatial gradient of the stress $S_i$ before events with $1 \le M \le 100$ and $80\le N \le 170$. Straight lines are power-law fits for $K < 0.1$; the obtained exponents are $1-2 H$ with $H$ the Hurst exponent associated to the large scales of the stress profiles. 
d: Exponents $\alpha$, $\beta$, $1+B$ and the prediction $1+(\beta-1)/\alpha$ as a function  of $K_2$. 
Symbols are best fits of the power-law exponents of fig 1a and 1b. 
Continuous lines are the predictions obtained from the fBm analogy ($\alpha=1+H$, $\beta=2-H$, $1+B=1+(1-H)/(1+H)$) with $H$ obtained from the large scales of the gradient of the stress, as in fig 1c.}
\label{figPNcoulomb}
\end{figure*}

We start with 1D geometry and  consider $N_t$ sliders on a line at positions $x_i$.  Each slider is connected to its nearest neighbors with a spring of stiffness  ${k}_2$ 
 and with a spring of stiffness ${k}_1$ to a plate that moves at constant velocity ${v}_0$.  The driving force on the i-th slider is
\begin{equation}
{S}_{i}=-{k}_2 (2 {x}_i- {x}_{i+1}-{x}_{i-1})+ {k}_1 ({v}_0 \, {t} - {x}_i)\,.
\label{defFi}
\end{equation}
When the sliders are all at rest, they experience a linear in time increasing load until  ${S}_{i}$ reaches the static friction force ${F}_s$ for a given  slider $i$, which starts to move with  velocity  ${v}_i$ and is then subject to the dynamic friction force $F_d$.  First, we consider the case of a constant, Coulomb-like, friction $F_d$, a model that has received little attention compared to the  standard Burridge-Knopoff (BK) model \cite{BK} that considers a velocity weakening behavior for $F_d$. The dynamical equation is then $m {\dot{v}}_i = {S}_{i}- {F}_d $. 
In addition, a slider is not allowed to move backward. 
We set $m=k_1=F_s=1$, $N_t=800$, $v_0=10^{-6}$, $F_d=0$ and  $k_2$ is varied between $5$ and $13$. 
The system alternates between a loading period with sliders at rest 
followed by a brief event initiated once one of the sliders starts moving
and can put into motion a varying number of sliders. These sudden events are  the EQ of the model. 
The system has a chaotic behavior with large fluctuations of the EQ moment $M=\sum \Delta x_i$ where $\Delta x_i$ is   the total slip of the i-th slider during the EQ.   The PDF  $P(M)$  displays a power-law behavior $P\propto M^{-1-B}$ with exponent $B$ that varies with $k_2$, see  fig. \ref{figPNcoulomb} a. 


We also observe a wide  distribution of the spatial extent of the event {\it i.e.} the number of masses involved in  each event $N$.   As for the  distributions of the moment, they display 
a power-law $P\propto N^{-\beta}$, see fig. 1b. In addition, the moment and the spatial extent are related. We calculate the value of $\langle M \rangle_{N}$ where the average is taken at fixed value of $N$.  At intermediate value of $N$ a power-law  $\langle M \rangle_{N} \propto N^{\alpha}$ is observed (not displayed here). 
$\alpha$ and $\beta$ both vary with $k_2$ and are displayed in fig. \ref{figPNcoulomb}d.


Using the rule of change of variable for a probability, together with $P(N)\propto N^{-\beta}$ and   $M \propto N^{\alpha}$, we obtain that the  distribution of $M$ is a power-law, $P(M)\propto M^{-1-B}$, and predict   
\begin{equation}
B=\frac{\beta-1}{\alpha} \,.
\label{relexpo}
\end{equation}
 This prediction is verified, see  fig.  \ref{figPNcoulomb}c. 


The exponent for the distribution of the moment is thus related to the one of the spatial extent and to the one of the moment vs spatial extent relation. To progress in the understanding of the $B-$value, we   need to understand what sets the value of $\alpha$ and $\beta$. 
It appears that the stress field $S_i$, as defined in eq. \ref{defFi}, plays a particularly important role. Indeed,  $S_i$ determines the slip of the event: its size and moment.  
More precisely, we observe   that for events that involve a sufficient number of masses, there exists a simple effective linear relation between $S_i$ and $\Delta x_i$. The slip profiles thus correspond to excursions above a fixed threshold of the stress field. In particular the length of the event is the length at which the stress profile returns to its initial value (the return time for a random walk) and the moment of the earthquake is the surface below the stress excursion. 


Predictions on the EQ properties can thus be obtained from the properties of the stress profile. We  show the power spectrum density of the spatial gradient of the stress profile in fig. \ref{figPNcoulomb}c. 
At small wavevector, $K<0.1$, the spectrum displays a power-law   $K^{1-2 H}$.  
We associate this behavior to  a fractional Brownian motion (fBm) of Hurst exponent H \cite{hurst}.   More precisely, the large scales of the stress profile  have properties similar to those of the excursions of a fBm. As the slip is proportional to the stress, we can use properties of the fBm to predict the ones of the EQ. 
Return times of a fBm are distributed as $P(N)=N^{H-2}$ \cite{returnhurst} so that the exponent for the EQ size is $\beta=2-H$. Typical excursions of a fBm of size $N$ are of size $N^H$ so that the area covered by a fBm excursion scales as $N^{1+H}$. This implies that the exponent of the moment-size relation  is $\alpha=1+H$. We  thus obtain a prediction for the B-value using  equation \ref{relexpo} as $1+B=1+\frac{1-H}{1+H}$. This prediction, together with $\alpha=1+H$ and $\beta=2-H$ are verified and are displayed as continuous lines in fig.\ref{figPNcoulomb}d, using $H$ fitted from the spectrum of the gradient of the stress profile. 

Let us  sum up on the behavior of the Coulomb BK model. The large scales of the  stress  behave as a fBm of Hurst exponent H. The EQ slips are proportional to excursions of this fBm. This analogy between fBm and earthquake properties allows us to predict the values of the exponents $\alpha$, $\beta$ and $B$ as a function of one  single parameter $H$, the Hurst exponent of the large scales of the stress.

We have successfully performed a similar investigation on other 1D  models such as a BK model with a slip-weakening friction force or the standard BK model \cite{BK} and could understand for each of these models the value of $B$.  As for the Coulomb friction model, all  these  systems can be described in a more simple manner by considering the coupled dynamics of two fields:  the stress before an event and the total slip during an event.  
In all these systems, the stress is scale invariant at large scale. This scale invariance is spontaneously built up during the evolution of the stress caused by the successive action of  EQ of varying size.  This mechanism, simple and robust, provides an explanation for the properties of EQ in models, and might also be at work in experiments and nature. 


\begin{figure}
a\includegraphics[width=8.5cm]{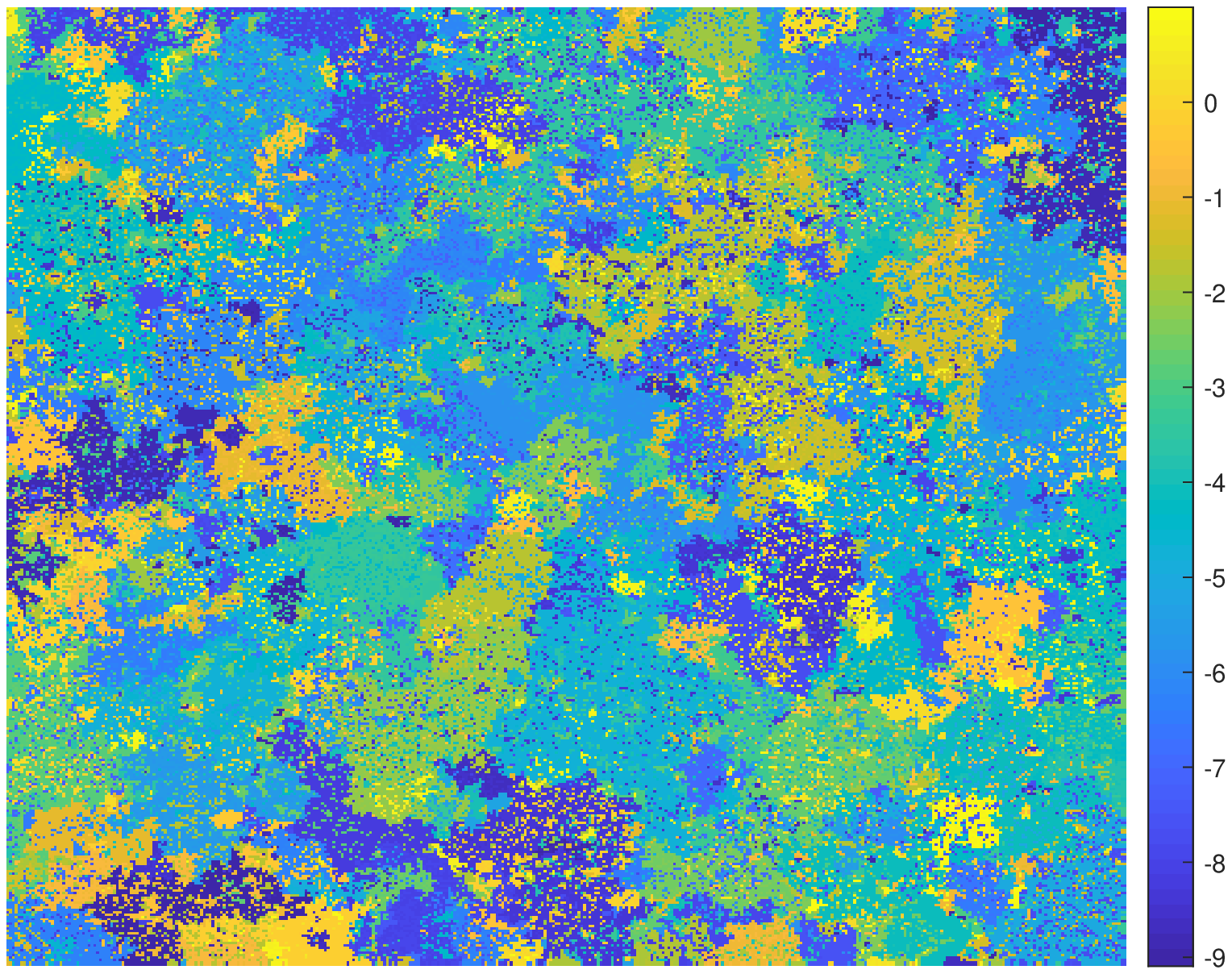}
b\includegraphics[width=8.5cm]{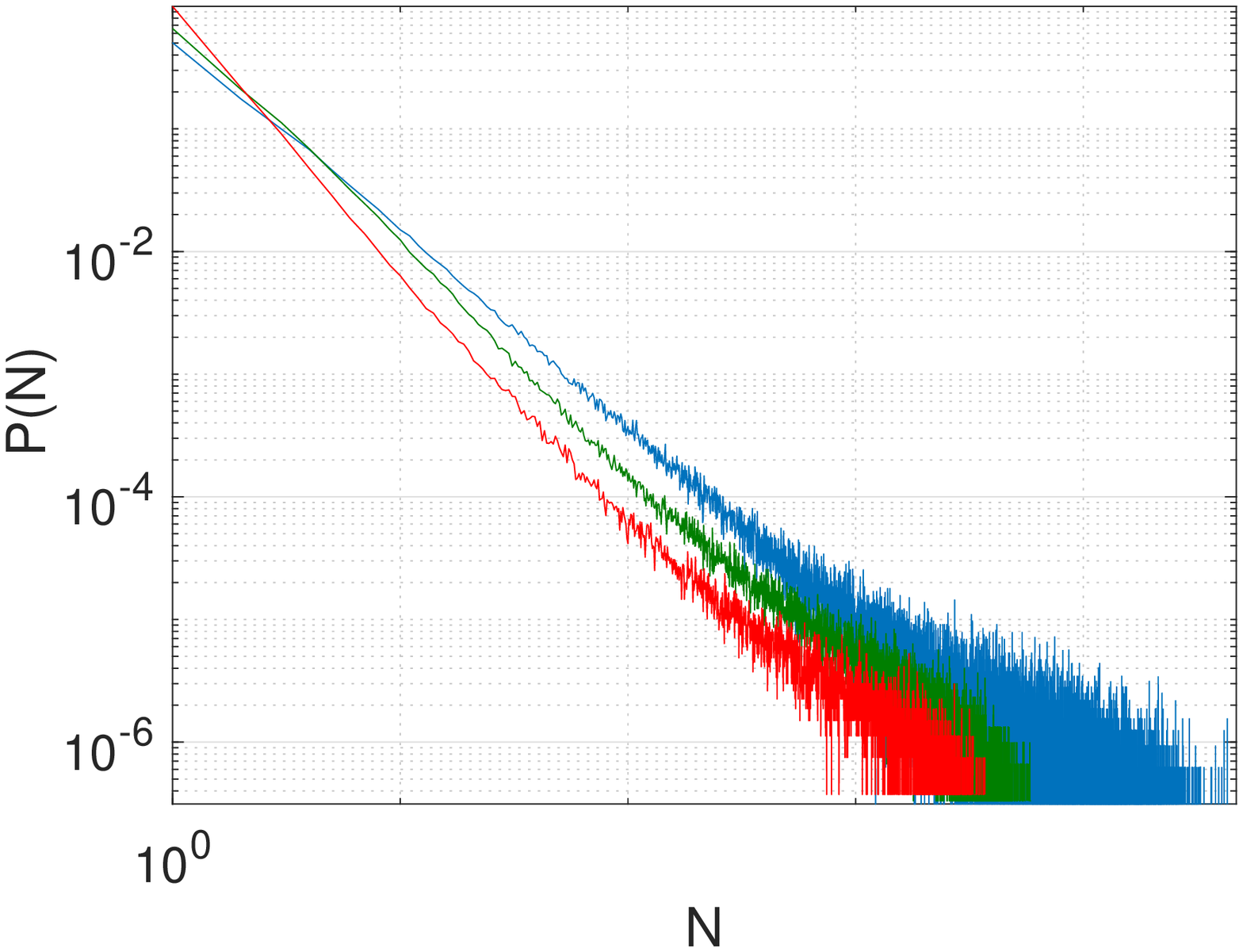}
c\includegraphics[width=8.5cm]{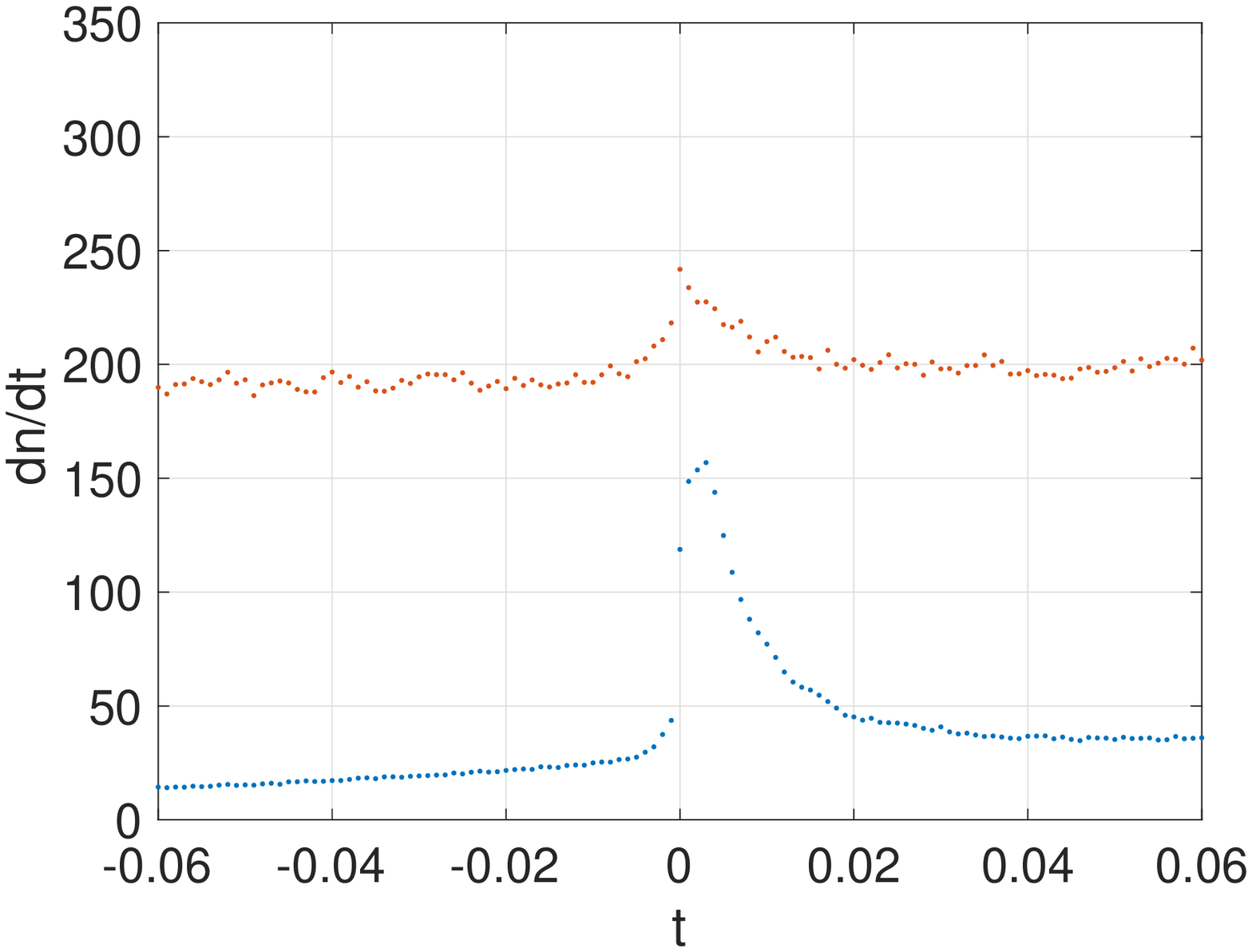}
\caption{For the 2D model with $D_1=1$, $D_2=10$, $D_3=0.1$, $N_t=400^2$.  a: Snapshot of  the stress field  as a function of position  for  $s=0$.  Values on the color bar on the right. b: PDF of EQ size $P(N)$ for red $s=0$, green $s=1$, blue $s=1.5$. c: 
 Averaged number of events per unit of time as a function of duration to a main event of size $N \ge 100$. Only events with hypocenter located  at a distance smaller than $50$ from the main shock are considered. Red: $s=0$ and blue: $s=1.5$.  }
\label{figBi}
\end{figure}

In the following, inspired by this mechanism, we introduce a 2D model based on such evolution rules for the stress field. This model is made as simple as possible and thus omits several features of real EQ. Yet we will observe that the stress field displays scale invariance and we will obtain additional predictions for the EQ properties in 2D.
We consider $N_t$ sites on a square lattice. Between events, the stress at each site $S_i$ increases linearly in time at a rate $v_o$. When the stress at one site, say $i_0$, reaches a threshold value, $S_c$, an earthquake is initiated. Let $D_1$ be a constant, we identify sites which stress is larger than  $S_c-D_1$. All  these sites can be classified into clusters made of neighboring sites,  see fig. \ref{figBi} for an example of stress field. 
Only the sites which belong to the same cluster as $i_0$ participate to the earthquake. 
Let $N$ be the number of sites in this cluster. As in nature \cite{valb}, we assume that their  motion during the earthquake is proportional to the  length of the earthquake, {\it i.e.} $\sqrt{N}$, which results in a moment $M=N^{3/2}$. 

After an event, the stresses of the moving sites are set to new values equal to their former value minus a stress drop $D_d$ equal to a constant $D_2$ plus a random term.  The system is then back into the initial stage for which the sites are at rest and the stress increases linearly in time.  
We consider a random term that can be spatially correlated. Its amplitude is $D_3$. It is self-affine with Hurst exponent $H_n=s-1$ where $s$ is a parameter 
 and is obtained as follows.  
 A random field is calculated with values of uniform probability between $0$ and $D_3$. The field is Fourier transformed in space, filtered by multiplication with a kernel $K^{-s}$ with $K$ the wavevector and then inversed Fourier transformed. This procedure generates a correlated random surface with Hurst exponent $H_n=s-1$ 
\cite{randsurf,reviewGFF}. The random field is calculated over the whole $N_t$ sites but only the values corresponding to the moving sites are used. To gain computational time, we perform this procedure only for events of size larger than $5$, otherwise an uncorrelated random field is taken with value uniformally sorted between $0$ and $D_3$.  
For $s=0$, the random stress drop is  an uncorrelated field (white noise), whereas it is correlated for positive $s$  \cite{reviewGFF}.  
In this model, the interaction between sites takes place when sites which belong  to the same cluster move and the value of the stress of these sites is reduced by a uniform value and a correlated one for $s\neq 0$. 


Numerical simulations are performed for $v_0=1$,  $S_c=1$, $D_1=1$ and $D_2=10$, $D_3=0.1$, $N_t=400^2$. 
 The distribution of size of the EQ is a power-law, see fig.\ref{figBi}, 
 with exponent $\beta$ that depends on $s$. For $s=0$,  a best-fit for $3 \le N \le 80$ leads to $\beta=2.046 \pm 0.015$. It decreases to $\beta=1.93 \pm 0.01$ for $s=1$ and  $\beta=1.63 \pm 0.01$ for $s=1.5$.  Results similar to the ones of $s=0$ are obtained for small positive $s=0.5$ or negative $s=-1$.  In these models, the relation between size and moment is taken to be $M=N^{3/2}$, so that the B-values ($B=2 (\beta-1)/3$) are  equal to  $0.697$, $0.62$, $0.42$ for $s=0$, $1$ and $1.5$ respectively.





In addition to these wide distributions of moments, aftershocks are observed, see  fig. \ref{figBi} c.  
Both an Omori's law after the mainshock and an inverse Omori's law of smaller amplitude before the mainshock are visible. 
The larger $s$, the stronger is the increase of activity in the vicinity of the mainshocks.
The data can be fitted using the standard Omori's law, $\frac{dn}{d\tau} \propto (\tau+a)^{-1}$ at least for $\tau$ not too small.  
We add that the distribution of interevent time \cite{corral} also shows the existence of clusterring when $s> 0$, a property that is related to Omori's law \cite{sornette,manip,manip2}. 

As for 1D geometry, the observed behavior can be understood by an analysis  of the spatial distribution of the stress. The stress field is a surface and events involve clusters of sites for which the stress is larger than a threshold value.  This problem is classical in statistical physics as it covers a variety of analogous situations \cite{ishenko}. 
If the comparison of the stress with a  threshold  defines whether a site is occupied or empty, then  the problem of percolation is recovered \cite{stauffer}. 
From this analogy, we expect the existence of a critical point close to which the cluster size are distributed following a power-law with exponent
$\tau_F$, so-called Fisher exponent. 
If the heights of the surface at different positions are uncorrelated, standard percolation takes place and $\tau_F=187/91$ \cite{stauffer}. If the heights are correlated as $\langle (h(x+R)-h(x))^2\rangle^{1/2}=R^{H} $, then correlated percolation  takes place.  For $H \le -3/4$, correlations do not affect the critical behavior that remains identical to the one of uncorrelated percolation.  For correlated surfaces with $H \ge -3/4$, the critical exponents vary with $H$ \cite{ishenko, zierenberg2018}.

 We have calculated the  Hurst exponent  of the stress field. At large scales,  we measure $H\simeq -0.32$ for $s=1.5$,    $H\simeq -0.5$ for  $s=1$,  and  $H\simeq -1$ for $s=0$. We point out that the Hurst exponent of the stress field varies with the one of the noise term ($H_n=s-1$) but they are different. Indeed, the stress field results from the successive addition of many random terms with varying size, so that the stress drop and the total stress have different properties. 


Considering that  the stress field has the same properties as  a random field with Hurst exponent $H$ that depends on $s$, and assuming that the system is in the vicinity of the percolation critical point, we expect that the distribution of size of clusters above a cut follows a power-law of exponent $\tau_F$. This is in agreement with the exponent of the distribution of the size of the event for small $s$ as we have measured $\beta=2.046 \pm 0.015$ for $s=0$, very close to Fisher prediction for uncorrelated percolation $\tau_f=187/91 \simeq 2.055$ . 
 Written in term of $b$-values, we  predict $b=96/91\simeq 1.05$ and $B=64/91\simeq 0.70$ for weakly correlated stress surfaces. If the stress  has strong correlations, namely $H>-3/4$, the analogy with correlated percolation predicts a decrease in the b-value. This is indeed observed in our model as $B=0.697$ for $H=-1$ ($s=0$) and similar values are obtained  for moderate correlations ($s=-1$ and $s=0.5$). Increasing the correlation and for $H>-3/4$, we  measure as expected a decrease in $B$ since  $B=0.62$ for $H= -0.5$ and $B=0.42$ for $H=-0.32$. 


\begin{figure}
\includegraphics[width=9cm]{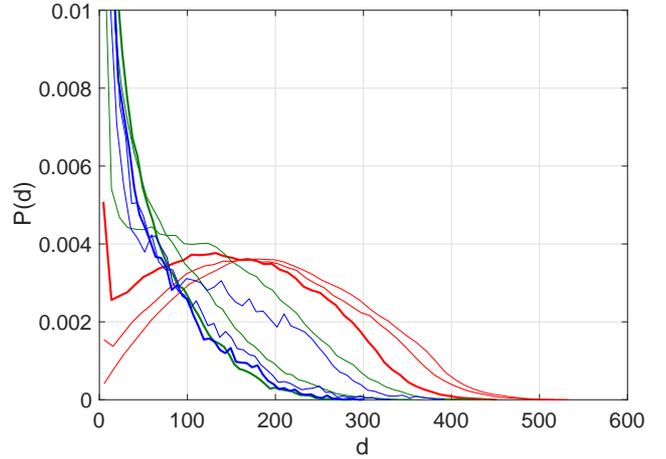}
\caption{
PDF of the distance $d$ between the largest cluster of occupied sites and the other clusters for a random surface with Hurst exponent $H=-1$ in red, $H=0$ in green and $H=0.5$ in blue. The fraction of  occupied sites  increases with the line thickness and is $0.3$, $0.5$ and $0.55$. The total number of sites is $400^2$. }
\label{figomori1}
\end{figure}

Aftershocks and foreshocks  also find a simple explanation based on the spatial structure of the stress field.
We have simulated random surfaces with various Hurst exponents $H$  by multiplication in Fourier space with a well-chosen power-law
using the method of \cite{randsurf}.  
For each surface, we identify the sites which value is above a given threshold. These sites are said to be occupied using the terminology of percolation. We calculate  $p$ the fraction of occupied sites. We then identify the clusters of connected sites and  calculate the distance $d$ between the largest cluster and the other clusters. The distribution of $d$, $P(d)$, is a characteristic of the spatial organization of the clusters. It is displayed in fig.\ref{figomori1} for different values of $p$ and $H$.  
A clear divergence of $P(d)$ for small $d$ is visible when $H$ gets larger. 
In other words, increasing $H$ leads to a localization of the clusters close to the largest cluster. 
Correlated random surfaces, with  $H>0$, thus display level sets  which positions are correlated. This property of random surface has never been described so far.   It can be translated in terms of the spatial distribution of large values of stress in a fault: for a correlated stress field, if a domain has a large value of stress, other clusters with large stress are likely to be located in its vicinity. 
Clusterring in space of individual clusters with large stress is thus a consequence of the geometrical properties of the stress field that behaves at large scale as a correlated random surface. This provides an explanation for the Omori's law: in the vicinity of a mainshock, there exist clusters which stress is large and close to initiate an EQ. This mechanism is purely geometrical and does not require additional phenomena (stress transfer, pore pressure dynamics, viscosity....).

In nature,  it has been reported that slip profiles, stress or friction fields have self-affine properties \cite{rough, renard}, in line with our description. 
In addition, the b-values that we predict, $b=96/91\simeq1.05$ for weakly correlated stress field and a smaller one for strong correlation,  are  compatible with most reported natural values \cite{valb,Takareview}. 
We also note that the prediction $B=64/91$ is in perfect agreement with the measured exponent in an experiment of  sheared granular matter that measures $B=0.71 \pm 0.01$ for the released energy during the events \cite{manip}. 
To finish on this brief comparison with  datas, we add that  \cite{Schorlemmer}, \cite{Scholz} describe a variation of $b$ with properties of the faults. Our work provides an explanation for these effects: the fault properties affect the self-affine behavior of the stress field or  the moment vs size relation. These properties  in turn modify the $b$-value.

Essentially, the GR law and Omori's law result from the scale invariance  of the stress. This scale invariance builds up during the iterative coupled evolution of the stress and the slip. Power-law distributions of the moment are then a consequence of this scale invariance and their exponent are related to the the statistical properties of random curves or surfaces. 

This approach is promising. First, it can be applied to other models of EQ or of similar effects such as avalanches in order to understand  the emergence of power law  distributions. 

Second, the analysis of the two steps of evolution  sheds the light on new problems in statistical mechanics. The  stress to slip problem is a non standard application of random curves: the stress profile, a  random trajectory,  defines through  possibly nonlinear rules the slip distribution which is also a random curve. In probability theory, this situation belongs to the problem of random polymers and our results open the way to the study of new classes of such models.  

Third, the second step in the modeling, from slip to stress field, is obviously crucial here. The stress field evolves at each event because  the slip decreases the stress of the moving sites but both the stress and the slip fields are  coupled as the slip is determined by the  stress profile before the event.   
As a consequence, EQ properties  result from the non trivial dynamical evolution of a random interface instead of resulting from the self-organization close to an equilibrium critical point \cite{soc}.  It remains to be understood how and when such evolving interfaces tend  towards a self-affine geometrical structure.   For these questions also, very little is known and we expect  that new classes of random interfaces will be identified in this context. The robustness of EQ-like behavior in models and nature pushes toward the existence of generic mechanisms able to generate these self-affine surfaces. 


F P gratefully acknowledges the Visiting Research Program in 2019 at the Earthquake
Research Institute, the University of Tokyo and the IEA action of CNRS. A S gratefully acknowledges the support of the European Research Council Grant REALISM (2016-Grant 681346). T H and T S are
also supported by the MEXT under ‘Exploratory Challenge on Post-K computer’ (Frontiers of Basic Science: Challenging the Limits). T H gratefully acknowledges additional
support from JSPS KAKENHI Grant JP16H06478. 



\begin{thebibliography}{4}

\bibitem{bookEQ} Scholz C. H., The Mechanism of Earthquakes and Faulting, Cambridge University Press, 2019.

\bibitem{valb} 
 H. Kanamori and E. E. Brodsky, Reports on Progress in Physics 67(8)  1429 (2004). 


\bibitem{Takareview} H. Kawamura et al., Rev. of modern physics, {\bf 84}, 839 (2012)

\bibitem{Omori} T. Utsu et al., J. Phys. Earth {\bf 43}, 1-33 (1995).


\bibitem{BK}  Burridge, R., and L. Knopoff, 1967, Bull. Seismol. Soc. Am. 57,
341.  Carlson, J. M., J. S. Langer, and B. E. Shaw, 1994, Rev. Mod. Phys.
66, 657.

\bibitem{OFC} Z. Olami, H. J. S. Feder and K. Christensen, Physical Review Letters, 68, 1244-1248 (1992).

\bibitem{PRE77}  J. Xia et al., Phys. Rev. E {\bf 77}, 031132 (2008).


\bibitem{hurst} 
B. Mandelbrot and J. W. Van Ness, SIAM  Review  {\bf  10},   422-437 (1968).

\bibitem{returnhurst} M. Ding and W. Yang, Phys. Rev. E {\bf 52}, 207 (1995). 






\bibitem{randsurf} C.P. de Castro et al., Scientific Report {\bf 7}, 1961 (2017).  


\bibitem{corral} A. Corral, Phys. Rev. Lett. {\bf 92}, 108501 (2004).

\bibitem{sornette} A. Saichev and D. Sornette, 
Phys. Rev. Lett. {\bf 97}, 078501 (2006).

\bibitem{manip} S. Lherminier et al., Phys. Rev. Lett. {\bf 122}, 218501 (2019). 

\bibitem{manip2} D. Houdou et al., Commun Earth Environ {\bf 2}, 90 (2021). 

\bibitem{randsurf} C.P. de Castro et al., Scientific Report {\bf 7}, 1961 (2017). 

\bibitem{reviewGFF} A. Lodhia et al., Fractional Gaussian fields: A survey, Probability Surveys, 13, 1-56 (2014).  

\bibitem{ishenko} M. B. Ishenko, Reviews of modern physics {\bf 64}, 961 (1992).

 


\bibitem{stauffer} D. Stauffer and A. Aharony, Introduction to percolation theory, Taylor and Francis (2010).  

\bibitem{zierenberg2018} J Zierenberg et al., Physical Review E 96 (6), 062125 (2017).



\bibitem{rough} F. Renard et al., GRL {\bf 40}, 83-87  (2013). S. Abe, and H. Deckert, Solid Earth Discuss., preprint 2021. L. Bruhat et al., Geophys. J. Intern., {\bf  220}, 1857–1877 (2020).

\bibitem{renard} T. Candela et al., Geophysical Journal International, {\bf 187}, 959-968  (2011).


\bibitem{Schorlemmer} D. Schorlemmer et al., Nature 437 (7058), 539-42 (2005).

\bibitem{Scholz} C.H. Scholz, Geophys. Res. Lett., 42,  1399–1402 (2015). 

\bibitem{soc} P. Bak, C. Tang, and K. Wiesenfeld, Phys. Rev. Lett. 59,
381 (1987).







\end{thebibliography}
\end{document}